\documentclass[hyper]{JHEP} % 10pt is ignored!

\usepackage{epsfig}
\newcommand\fverb{\setbox\pippobox=\hbox\bgroup\verb}

\newcommand\fverbdo{\egroup\medskip\noindent%

            \fbox{\unhbox\pippobox}\ }

\newcommand\fverbit{\egroup\item[\fbox{\unhbox\pippobox}]}

\newbox\pippobox

\title{Note About Hamiltonian Formalism for Newton-Cartan
String and $p-$Brane}
\author{J. Kluso\v{n}\\
Department of
Theoretical Physics and Astrophysics\\
Faculty of Science, Masaryk University\\
Kotl\'{a}\v{r}sk\'{a} 2, 611 37, Brno\\
Czech Republic\\
E-mail: \email{klu@physics.muni.cz}} \preprint{}

 \abstract{We construct non-relativistic string and p-brane actions
 in Newton-Cartan background using the limiting procedure from the
relativistic string and p-brane action in general background. We also 
find their Hamiltonian formulations when however we restrict ourselves
to the case of the vanishing gauge field $m_\mu$.}
\def\bA{\mathbf{A}}

\def\ttau{\tilde{\tau}}

\def\be{\begin{equation}}

\def\ee{\end{equation}}

\def\bea{\begin{eqnarray}}
\def\bh{\bar{h}}
\def\eea{\end{eqnarray}}

\def\mH{\mathcal{H}}

\newcommand{\mG}{\mathcal{G}}

\def \bA{\mathbf{A}}

\newcommand{\bT}{\mathbf{T}}

\newcommand{\ba}{\mathbf{a}}

\newcommand{\mL}{\mathcal{L}}

\def\bai{(\mathbf{a}^{-1})}

\def\pb #1{\left\{#1\right\}}

\begin{document}
%%%%%%%%%%%%%%%%%%%%%
%%%%Introduction %%%%%%%%%
%%%%%%%%%%%%%%%%%%%%
\section{Introduction }
Today it is well known that strong correlated systems 
in condensed matter can be sucessfully described with the help of non-relativistic holography, for review see for example \cite{Hartnoll:2016apf}. This duality is based on the idea that the strongly coupled theory on the boundary 
can be described by string theory in the bulk. Further, when the curvature of the space-time is small we can use the 
classical gravity instead full string theory machinery. In case of non-relativistic holography the situation is even more interesting since we have basically two possibilities: Either we  use Einstein metric with non-relativistic isometries
\cite{Son:2008ye,Balasubramanian:2008dm,Herzog:2008wg} or we 
introduce non-relativivistic gravities in the bulk 
\cite{Son:2013rqa,Janiszewski:2012nb}, like Newton-Cartan gravity \cite{Cartan:1923zea} 
\footnote{For some recent works, see 
	\cite{Bergshoeff:2017dqq,Bergshoeff:2016lwr,Hartong:2016yrf,Hartong:2015zia,
		Afshar:2015aku,	Bergshoeff:2015ija,Bergshoeff:2015uaa,Bergshoeff:2014uea,
		Andringa:2010it}.}
or Ho\v{r}ava gravity \cite{Horava:2009uw}. Then it is certainly very interesting
to study matter coupled to non-relativistic gravity. We can either study 
field theories on non-relativistic background as in 
\cite{Bergshoeff:2015sic,Bagchi:2015qcw,Festuccia:2016caf,Geracie:2015dea,Jensen:2014aia,Hartong:2014pma} or particles \cite{Kuchar:1980tw,Barducci:2017mse,Bergshoeff:2014gja,Kluson:2017pzr}
or even higher dimensional objects, as for example non-relativistic strings
and p-branes \cite{Andringa:2012uz,Harmark:2017rpg}. 

In this work we would like to focus on the  canonical formulation of non-relativistic
string and p-brane in Newton-Cartan background. The starting point of our analysis is 
the relativistic string in general background that couples to NSNS two form.
Then we use the limiting procedure that was proposed in 
\cite{Bergshoeff:2015sic} and try to find corresponding string action. 
Note that this is different limiting procedure than in case of the non-relativistic string in flat background where the non-relativistic limit is performed 
on coordinates \cite{Gomis:2000bd,Gomis:2005bj,Gomis:2005pg}
\footnote{For recent work, see for example \cite{Batlle:2016iel,Kluson:2017djw,Kluson:2017ufb,Kluson:2017vwp}.}.
It is important to stress that if we apply this limiting procedure that leads to corank-1 spatial metric and rank one temporal metric of Newton-Cartan gravity to the case of the string action 
 we find that there is no way how to ensure that this action is finite. In order
 to resolve this problem we have to select two flat target space   longitudial directions exactly in the same way as in \cite{Andringa:2012uz}. Then we  propose such an ansatz for NSNS two form field that is constructed with 
 the help of the fields that define Newton-Cartan geometry and where the divergent contribution from the coupling to NSNS two form exactly cancels the divergent
 contribution coming from Nambu-Gotto part of the action. As a result we obtain 
 an action for the string in Newton-Cartan background that was proposed in 
 \cite{Andringa:2012uz} using different procedure. 
 As the next step we proceed to the canonical formulation of this theory. Then however
 we encounter an obstacle in the form that we are not able to invert relation 
 between conjugate momenta and velocity in case of non-zero gauge field $m_\mu^{ \ a}$ whose explicit definition will be given in the next section. For that reason 
 we restrict ourselves to the case of the zero gauge field keeping in mind that
 the case of on-zero gauge field deserves further study. Then we find  Hamiltonian for this non-relativistic string that is linear combination of two first class constraints
 which is the manifestation of the fact that two dimensional string action 
 is invariant under world-sheet diffeomorphism.  As the next step  we generalize this analysis to the case of $p-$brane. 
 We firstly determine well defined action for non-relativistic p-brane 
 when we consider specific form of the background $p+1$ form that couples to the world-volume of p-brane. Then we introduce an equivalent form of $p-$brane action 
 that allows us to consider canonical analysis of this theory. Finally we determine constraint structure of this theory and we show that there are $p+1$ first class constraints, $p-$ spatial diffeomorphism constraints and one Hamiltonian constraint. We again show that these constraints are the first class constraints. 
 
 This paper is organized as follows. In the next section (\ref{second}) we determine the form of non-relativistic string in Newton-Cartan background and perform its Hamiltonian analysis. Then in section (\ref{third}) we generalize this analysis to the case of 
 $p-$brane. Finally in conclusion (\ref{fourth}) we outline our results and suggest possible extension of this work.

\section{Review of the Non-relativistic Limit for Nambu-Gotto String}
\label{second}
In this section we derive non-relativistic form of the string action 
in Newton-Cartan background using the limiting procedure developed in 
\cite{Bergshoeff:2015uaa}. We start with 
the Nambu-Gotto form of the action in the general background
\begin{equation}\label{funstringact}
S=-\ttau_F \int d\tau d\sigma\sqrt{-\det (E_\mu^{ \ A}E_\nu^{ \ B}\eta_{AB}
	\partial_\alpha x^\mu\partial_\beta x^\nu)}+\ttau
\int d\tau d\sigma B_{\mu\nu}\partial_\tau x^\mu \partial_\sigma x^\nu \ , 
\end{equation}
where $E_\mu^{ \ A}$ is $d-$dimensional vierbein so that the metric components have the form 
\begin{equation}
G_{\mu\nu}=E_\mu^{ \ A}E_\nu^{ \ B}\eta_{AB} \ , \eta_{AB}=\mathrm{diag}(-1,1,\dots,1)
\end{equation}
Note that the metric inverse $G^{\mu\nu}$ is defined with the help of the inverse vierbein $E^\mu_{ \ B}$ that obeys the relation 
\begin{equation}
E_\mu^{ \ A}E^\mu_{ \ B}=\delta^A_{ \ B} \ ,  \quad E_\mu^{ \ A}E^\nu_{ \ A}=
\delta_\mu^{ \ \nu} \ .
\end{equation}
Further, $B_{\mu\nu}$ is NSNS two form field that plays the crucial role in the
limiting procedure. Finally $x^\mu \ ,\mu=0,\dots,d-1$ are embedding coordinates of the string where the two dimensional world-sheet is parameterised by  $\sigma^\alpha\equiv(\tau,\sigma)$. 

Let us now proceed to the brief description of the procedure that 
leads to Newton-Cartan background from general background, for more detailed discussion, see the original paper \cite{Bergshoeff:2015uaa}. The starting point
is following ansatz for $d-$dimensional vierbein \cite{Bergshoeff:2015uaa}
\begin{equation}\label{vierbeinomega}
E_\mu^{ \ 0}=\omega\tau_\mu+\frac{1}{2\omega}m_\mu \ , \quad  E_\mu^{ \ a'}= e_\mu^{ \ a'} \ , 
\end{equation}
where $a'=1,\dots,d-1$ and where $\omega$ is free parameter which goes to infinity in the Newton-Cartan limit. Note that in this case the metric has the form 
\begin{eqnarray}
& &G_{\mu\nu}=E_\mu^{ \ A}E_\nu^{ \ B}\eta_{AB}
=-\omega^2 \tau_\mu \tau_\nu-\frac{1}{2}\tau_\mu m_\nu-\frac{1}{2}\tau_\nu m_\mu+h_{\mu\nu}+\frac{1}{4\omega^2}m_\mu m_\nu=\nonumber \\
&=&-\omega^2\tau_\mu\tau_\nu
+\bh_{\mu\nu}+\frac{1}{4\omega^2}m_\mu m_\nu \ , \quad 
\bh_{\mu\nu}=h_{\mu\nu}-\frac{1}{2}\tau_\mu m_\nu-\frac{1}{2}\tau_\nu m_\mu \ , \quad 
h_{\mu\nu}=e_\mu^{ \ a'}e_\nu^{ \ b'}\delta_{a'b'} \ . \nonumber \\
\end{eqnarray}
Inserting this metric into the Nambu-Gotto action and performing expansion 
with respect to $\omega$ we obtain 
\begin{equation}
S=-\ttau_F \omega^2\int d\tau d\sigma
\sqrt{-\det\ba} -\frac{\ttau_F}{2}\int d\tau d\sigma 
\sqrt{-\det\ba}\ba^{\alpha\beta}\bh_{\alpha\beta} \ , 
\end{equation}
where we defined 
\begin{equation}\label{actnaive}
\ba_{\alpha\beta}=\tau_{\mu\nu}\partial_\alpha x^\mu \partial_\beta x^\nu \ ,
\quad \ba^{\alpha\beta}\ba_{\beta\gamma}=\delta^\alpha_\gamma \ , \quad 
\bh_{\alpha\beta}=\bh_{\mu\nu}\partial_\alpha x^\mu \partial_\beta x^\nu \ . 
\end{equation}
We also used the fact that $\ba_{\alpha\beta}$ is $2\times 2$ matrix that is non-singular. 
Apparently we see from (\ref{actnaive}) that there is a term proportional to $\omega^2$ that cannot be made
finite by rescaling of $\ttau_F$. In case of the string in the flat non-relativistic
background such a term  
can be canceled with the suitable form of the background $NSNS$ two form. Further, 
, this two form field should be build from the fields  that define  Newton-Cartan theory as $m_\mu,\tau_\nu$. However it turns out that it is not possible to find such a  NSNS two form due to the fact that it has to be antisymmetryc in space-time indicies. In order to find solution of this problem we implement  the generalization of the Newton-Cartan gravity that was introduced in 
\cite{Andringa:2012uz}. Explicitly,  we split the target-space indices $A$ into $A=(a',a)$ where now $a=0,1$ label longitudial  and $a'=2,\dots,d-1$ correspond to transverse directions. Then we introduce $\tau_\mu^{ \ a}$ so that we write 
\begin{equation}
  \tau_{\mu\nu}=\tau_\mu^{ \ a}\tau_\nu^{ \ b}
\eta_{ab} \ , \quad  a,b=0,1 \ , \quad \eta_{ab}=\mathrm{diag}(-1,1) \ . 
\end{equation}
In the same way we introduce vierbein $e_\mu^{ \ a'}, a'=2,\dots,d-1$ and also  we generalize $m_\mu$ into $m_\mu^{ \ a}$. The $\tau_\mu^{ \ a}$ can be interpreted as the gauge fields of the longitudinal translations while $e_\mu^{ \ a'}$  as the gauge fields of the transverse translations 
\cite{Andringa:2012uz}. Then we can also introduce their inverses with respect to their longitudinal and transverse subspaces
\begin{eqnarray}
e_\mu^{ \ a'}e^\mu_{ \ b'}=\delta^{a'}_{b'} \ ,  \quad
e_\mu^{ \ a'}e^\nu_{ \ a'}=\delta_\mu^\nu-\tau_\mu^{ \ a}
\tau^\nu_{ \ a} \ , \quad \tau^\mu_{ \ a}\tau_\mu^{ \ b}=\delta_a^b \ , \quad 
\tau^\mu_{ \ a}e_\mu^{ \ a'}=0 \ , \quad 
\tau_\mu^{ \ a}e^\mu_{ \ a'}=0 \ . \nonumber \\
\end{eqnarray}
Performing this generalization implies following form 
of the vierbein 
\begin{equation}
E_\mu^{ \ a}=\omega \tau_\mu^{ \ a}+\frac{1}{2\omega}m_\mu^{ \ a} \ , \quad 
E_\mu^{ \ a'} =e_\mu^{ \ a'}
\end{equation}
so that we find following form of the metric
\begin{eqnarray}
G_{\mu\nu}&=&E_\mu^{ \ a}E_\nu^{ \ b}\eta_{ab}+E_\mu^{ \ a'}E_\nu^{ \ b'}\delta_{a'b'}
=\nonumber \\
&=&\omega^2 \tau_{\mu\nu}+h_{\mu\nu}+\frac{1}{2}\tau_\mu^{ \ a}m_\nu^{ \ b}\eta_{ab}+
\frac{1}{2}m_\mu^{ \ a}\tau_\nu^{ \ b}\eta_{ab}+\frac{1}{4\omega^2}m_\mu^{ \ a}m_\nu^{ \ b}
\eta_{ab} \ .  \nonumber \\ 
\end{eqnarray}
It was shown  in \cite{Bergshoeff:2015uaa} that in order to find the right form 
of the particle action in Newton-Cartan background we should consider following
ansatz for the background gauge field $A_\mu=\omega \tau_\mu-\frac{1}{2\omega}m_\mu$. 
In order to find correct form of the action for the string in Newton-Cartan background
we propose analogue form of the NSNS two form 
\begin{eqnarray}
B_{\mu\nu}&=&\left(\omega\tau_\mu^{ \ a}-\frac{1}{2\omega}m_\mu^{ \ a}\right)\left( \tau_\nu^{ \ b}-\frac{1}{2\omega}m_\nu^{ \ b}\right)\epsilon_{ab}=
\nonumber \\
&=&\omega^2\tau_\mu^{ \ a}\tau_\nu^{ \ b}\epsilon_{ab}-
\frac{1}{2\omega}(m_\mu^{ \ a}\tau_{\nu}^{ \ b}+
\tau_\mu^{\  a}m_\nu^{ \ b})\epsilon_{ab}+\frac{1}{4\omega^2}
m_\mu^{ \ a}m_\mu^{ \ b}\epsilon_{ab} \ , \quad    
  \epsilon_{ab}=-\epsilon_{ba} \ , \quad  
  \epsilon_{01}=1 \ . 
\nonumber \\ 
\end{eqnarray}
It is important that terms proportional to $\omega^{-2}$ and $\omega^{-4}$ vanish 
in the limit $\omega\rightarrow \infty$ while the divergent contribution cancel the divergence coming from Nambu-Gotto part of the action since
\begin{eqnarray}
&- &\ttau_F\omega^2\int d^2\sigma\sqrt{-\det\ba}+\frac{\ttau_F}{2} \int d^2\sigma
\epsilon^{\alpha\beta}B_{\mu\nu}\partial_\alpha x^\mu 
\partial_\beta x^\nu \nonumber \\
&=&
-\ttau_F \omega^2\int d^2\sigma \det \tau_\alpha^{ \ a}+
\omega^2 \frac{\ttau_F}{2}\int d^2\sigma \epsilon^{\alpha\beta}\epsilon_{ab}\tau_\alpha^{ \ a}\tau_\beta^{ \ b}=0 \ , 
\nonumber \\
\end{eqnarray}
where we introduced $2\times 2$ matrix $\tau_\alpha^{ \ a}\equiv \tau_\mu^{ \ a}\partial_\alpha x^\mu$ and where we used the fact that $\det \tau_\alpha^{ \ a}=
\frac{1}{2}\epsilon^{\alpha\beta}\epsilon_{ab}\tau_\alpha^{ \ a}\tau_\beta^{ \ b}$ where $\epsilon^{\alpha\beta}=-\epsilon^{\beta\alpha} \ , \epsilon^{01}=1
$ is antisymmetric symbol with upper indices. 

In summary we obtain the action for non-relativistic string in Newton-Cartan background in the form
\begin{equation}\label{nonrelstringNC}
S=-\frac{\tau_F}{2}\int d^2\sigma \sqrt{-\det\ba}\ba^{\alpha\beta}
\bh_{\alpha\beta} \ , 
\end{equation}
where $\ttau_F=\tau_F$. Note that the action (\ref{nonrelstringNC})
was derived previously in \cite{Andringa:2012uz}
using slightly different procedure. 

Our goal is to find Hamiltonian formulation of this theory. To do this we 
rewrite the Lagrangian density introduced in  (\ref{nonrelstringNC}) into the form 
\begin{eqnarray}\label{actionextended}
\mL&=&\frac{1}{4\lambda^\tau}(\bh_{\tau\tau}-2\lambda^\sigma
\bh_{\tau\sigma}+(\lambda^\sigma)^2\bh_{\sigma\sigma})-\lambda^\tau
\tau_F^2\bh_{\sigma\sigma}+\nonumber \\
&+&B^\tau \left(\lambda^\tau-\frac{\sqrt{-\det\ba}}{2\tau_F\ba_{\sigma\sigma}}\right)
+B^\sigma \left(\lambda^\sigma-\frac{\ba_{\tau\sigma}}{\ba_{\sigma\sigma}}\right) \ . 
\nonumber \\
\end{eqnarray}
It is easy to see an equivalence of these two Lagrangians since the 
 equations of motion for $B^\tau$ and $B^\sigma$
give
\begin{equation}
\lambda^\tau=\frac{\sqrt{-\det\ba}}{2\tau_F\ba_{\sigma\sigma}} \ , \quad 
\lambda^\sigma=\frac{\ba_{\tau\sigma}}{\ba_{\sigma\sigma}} \ . 
\end{equation}
Inserting this result into (\ref{actionextended}) and using the fact that 
\begin{eqnarray}
& &\frac{1}{\lambda^\tau}=
%-2\tau_F\frac{\ba_{\sigma\sigma}}{\det\ba}\sqrt{-\det\ba}=
-2\tau_F \ba^{\sigma\sigma}\sqrt{-\det\ba} \ , \quad 
\frac{\lambda^\sigma}{\lambda^\tau}
%=2\tau_F
%\frac{\ba_{\sigma\tau}}{\det\ba}\sqrt{-\det\ba}
=2\tau_F \ba^{\tau\sigma}
\sqrt{-\det\ba} \ , 
\nonumber \\
& & \frac{1}{4\lambda^\tau}
((\lambda^\sigma)^2-4\tau_F^2(\lambda^\tau)^2)=
-\frac{\tau_F}{2}\ba^{\sigma\sigma}\sqrt{-\det\ba}
\nonumber \\
\end{eqnarray}
we find that (\ref{actionextended}) reduces into (\ref{nonrelstringNC}).
Then  from (\ref{actionextended}) we obtain conjugate momenta
\begin{eqnarray}\label{pmu}
p_\mu&=&\frac{1}{2\lambda^\tau}\bh_{\mu\nu}\partial_\tau x^\nu-
\frac{\lambda^\sigma}{2\lambda^\tau}\bh_{\mu\nu}\partial_\sigma x^\nu
-B^\tau\frac{1}{2\tau_F\ba_{\sigma\sigma}}\tau_{\mu\nu}\partial_\alpha x^\nu \ba^{\alpha\tau}\sqrt{-\det\ba}-\frac{B^\sigma}{\ba_{\sigma\sigma}} \tau_{\mu\nu}\partial_\sigma x^\nu \ , \nonumber \\
P^\tau_B&=&\frac{\partial L}{\partial \partial_\tau B^\tau}\approx 0 \ , \quad 
P^\sigma_B=\frac{\partial L}{\partial\partial_\tau B^\sigma}\approx 0 
\ , \quad  P_\lambda^\tau=\frac{\partial L}{\partial \partial_\tau \lambda^\tau}\approx 0 \ , \quad  P_\lambda^\sigma=\frac{\partial L}{\partial \partial_\tau \lambda^\sigma}\approx 0 
\nonumber \\
\end{eqnarray}
Now we come to the most important problem in  our analysis which is an imposibility to invert the relation between $p_\mu$ and $\partial_\tau x^\mu$  in order to express $\partial_\tau x^\mu$ using the canonical variables. The reason why we are not able to do it is in the presence of the vector field $m_\mu^{ \ a}$ so that the contraction of the metric  $\bh_{\mu\nu}$  with $\tau^\mu$ is non-zero. For that reason we restrict to the simpler case when $m_\mu^{\  a }=0$. Despite of this simplification  we will see that even in this case the Hamiltonian formulation of the non-relativistic string in Newton-Cartan background is non-trivial task.
In case when $m_\mu^{ \ a}=0$ we have 
 $\bh_{\mu\nu}=h_{\mu\nu}$ and $\bh_{\mu\nu}h^{\nu\rho}
\bh_{\rho\sigma}=h_{\rho\sigma}$ and hence from (\ref{pmu}) we obtain
\begin{eqnarray}
\left(p_\mu+\frac{\lambda^\sigma}{2\lambda^\tau}h_{\mu\rho}
\partial_\sigma x^\rho\right) h^{\mu\nu}\left(p_\nu
+\frac{\lambda^\sigma}{2\lambda^\tau}h_{\nu\sigma}
\partial_\sigma x^\sigma\right)=\frac{1}{4(\lambda^\tau)^2}
\partial_\tau x^\mu h_{\mu\nu}\partial_\tau x^\nu \ . 
\end{eqnarray}
On the other hand let us multiply both sides of expression 
(\ref{pmu}) with  
$\tau^\mu_{ \ a}\eta^{ab}\epsilon_{bc}\tau_\rho^{ \ c}\partial_\sigma x^\rho$ and we obtain 
\begin{eqnarray}\label{pmumilt}
& &p_\mu \tau^\mu_{ \ a}\eta^{ab}\epsilon_{bc}\tau_\rho^{ \ c}\partial_\sigma x^\rho
=-B^\tau\frac{1}{2\tau_F \ba_{\sigma\sigma}}\partial_\alpha x^\mu \tau_\mu^{ \ a}\epsilon_{ab}
\tau_\rho^{ \ b}\partial_\sigma x^\rho\ba^{\alpha\tau}\sqrt{-\det\ba}-B^\sigma\frac{1}{\ba_{\sigma\sigma}}
\partial_\sigma x^\mu \tau_\mu^{ \ a}\epsilon_{ab}\tau_\nu^{ \ b}\partial_\sigma x^\nu\nonumber \\
\nonumber \\
&=&-B^\tau\frac{1}{2\tau_F\ba_{\sigma\sigma}}\partial_\tau x^\mu \tau_\mu^{ \ a}\epsilon_{ab}\tau_\rho^{ \ b}\partial_\sigma x^\rho
\frac{\ba_{\sigma\sigma}}{\det\ba}\sqrt{-\det\ba}=\frac{1}{2\tau_F}B^\tau\nonumber \\
\end{eqnarray}
using
\begin{equation}
\tau^\mu_{ \ a}\tau_{\mu\nu}=\tau_\nu^{ \ b}\eta_{ba}
\end{equation}
and also we used the fact that 
\begin{equation}
\sqrt{-\det \ba}=\det \tau_\alpha^{ \ a}=
\tau_\tau^a\tau_\sigma^b\epsilon_{ab} \ , 
\end{equation}
where $\tau_\alpha^{ \ a}=\tau_\mu^{ \ a}\partial_\alpha x^\mu$. 
%\begin{eqnarray}
%\ba_{\alpha\beta}=\partial_\alpha X^\mu \tau_\mu^{ \ a}\partial_\beta
%X^\nu \tau_\nu^{ \ b}\eta_{ab}=\bA_\alpha^{ \ a}\eta_{ab}\bA_\beta^{ \ b} \ , \nonumber \\
%\bA_{\alpha}^{ \ a}=\left(\begin{array}{cc}
%\partial_\tau X^\mu \tau_\mu^{  \ 0} & 
%\partial_\tau X^\mu\tau_\mu^{ \ 1} \\
%\partial_\sigma X^\mu \tau_\mu^{ \ 0} & 
%\partial_\sigma X^\mu \tau_\mu^{  \ 1} \\ \end{array}\right)\equiv
%\left(\begin{array}{cc}
%\tau_\tau^{ \ 0} & \tau_\tau^{ \ 1} \\ 
%\tau_\sigma^{ \ 0} & \tau_\sigma^{ \ 1} \end{array}\right)
%\end{eqnarray}
%As a check note that we can write 
%\begin{equation}
%\ba_{\alpha\beta}=\bA_\alpha^{ \ a}\eta_{ab}(\bA^T)^b_{ \ \beta}=
%\left(\begin{array}{cc}
%\tau_\tau^{ \ 0} & \tau_\tau^{ \ 1} \\ 
%\tau_\sigma^{ \ 0} & \tau_\sigma^{ \ 1} \end{array}\right)
%\left(\begin{array}{cc}
%-1 & 0 \\
%0 & 1 \\ \end{array}\right)
%\left(\begin{array}{cc}
%\tau_\tau^{ \ 0} & \tau_\sigma^{ \ 0} \\ 
%\tau_\tau^{ \ 1} & \tau_\sigma^{ \ 1} \end{array}\right)
%=\left(\begin{array}{cc} 
%\tau_\tau^{ \ a}\tau_\tau^{ b}\eta_{ab} &
%\tau_\tau^{ \ a}\tau_\sigma^{ b}\eta_{ab} \\
%\tau_\sigma^{ \ a}\tau_\tau^{ b}\eta_{ab} &
%\tau_\sigma^{ \ a}\tau_\sigma^{ b}\eta_{ab} \\ \end{array}\right)
%\end{equation}
%so that the determinant is equal to
%\begin{equation}
%\det \bA_{\alpha}^{ \ a}=\tau_\tau^{ \ 0}\tau_\sigma^{ \ 1}-\tau_\tau^{ \ 1}
%\tau_\sigma^{ \ 0}=\tau_\tau^a\tau_\sigma^be_{ab}
%\end{equation}
%that however also implies
%\begin{equation}
%\sqrt{-\det\ba}=\sqrt{\det \bA \det(-\eta)\det (\bA^T)}=
%\det \bA 
%\end{equation}
Then the relation (\ref{pmumilt})  implies following primary constraint
\begin{equation}\label{prim1}
\Gamma^\tau\equiv 
2\tau_Fp_\mu \tau^\mu_{ \ a}\eta^{ab}\epsilon_{bc}\tau_\rho^{ \ c}\partial_\sigma x^\rho-B^\tau \approx 0 \ . 
\end{equation}
On the other hand if we multiply the  relation (\ref{pmu}) with $\tau^{\mu\nu}\tau_{\nu\rho}\partial_\sigma x^\rho$ we obtain 
\begin{eqnarray}
p_\mu \tau^{\mu\nu}\tau_{\nu\rho}\partial_\sigma x^\rho
%B^\tau \frac{1}{2\tau_F \ba_{\sigma\sigma}}
%\partial_\sigma x^\mu \tau_{\mu\nu}\partial_\alpha x^\nu 
%\ba^{\nu\tau}\sqrt{-\det\ba}-B^\sigma \frac{1}{\ba_{\sigma\sigma}}
%\partial_\sigma x^\mu \tau_{\mu\nu}\partial_\sigma x^\nu=
%\nonumber \\
=-
B^\tau \frac{1}{2\tau_F \ba_{\sigma\sigma}}
\ba_{\sigma\alpha} 
\ba^{\alpha\tau}\sqrt{-\det\ba}-B^\sigma =
-B^\sigma \  
\nonumber \\
\end{eqnarray}
and hence we obtain second primary constraint
\begin{equation}
\Gamma^\sigma\equiv p_\mu \tau^{\mu\nu}\tau_{\nu\rho}\partial_\sigma x^\rho+B^\sigma\approx 0 \ . 
\end{equation}
As a result  the extended
 Hamiltonian with all primary constraints included has the form 
\begin{eqnarray}
H_E&=&\int d\sigma\left(\lambda^\tau (p_\mu h^{\mu\nu}p_\nu+\tau_F^2 h_{\sigma\sigma})
 -B^\tau \lambda^\tau-B^\sigma \lambda^\sigma
+\lambda^\sigma p_\mu h^{\mu\nu}h_{\nu\rho}\partial_\sigma x^\rho+\right.
\nonumber \\
&+&\left. U_\tau \Gamma^\tau+U_\sigma \Gamma^\sigma +v
_\tau^B P^\tau_B+v_\sigma^B P_B^\sigma+v^\lambda_\tau P_\lambda^\tau+
v_\sigma^\lambda P_\lambda^\sigma\right) \ . 
\nonumber\\
\end{eqnarray}
Let us now analyze the requirement of the preservation of the primary constraints 
$P_\lambda^\tau\approx 0 , P_\lambda^\sigma\approx 0$. In case of $P_\lambda^\tau$ we obtain 
\begin{eqnarray}
\partial_\tau P_\lambda^\tau&=&\pb{P_\lambda^\tau,H_E}=
-p_\mu h^{\mu\nu}p_\nu-\tau_F^2 h_{\sigma\sigma}+B^\tau=
\nonumber \\
&=&
-p_\mu h^{\mu\nu}p_\nu-\tau_F^2h_{\sigma\sigma}+2\tau_F p_\mu
\tau^\mu_{ \ a}\eta^{ab}\epsilon_{bc}\tau_\rho^{ \ c}\partial_\sigma x^\rho-
2\tau_F \Gamma^\tau \approx \nonumber \\ 
&\approx &-p_\mu h^{\mu\nu}p_\nu-\tau_F^2h_{\sigma\sigma}+2\tau_F p_\mu
\tau^\mu_{ \ a}\eta^{ab}\epsilon_{bc}\tau_\rho^{ \ c}\partial_\sigma x^\rho\equiv -\mH_\tau
\approx 0 \nonumber \\
\end{eqnarray}
and also 
\begin{equation}
\partial_\tau P_\lambda^\sigma=\pb{P_\lambda^\sigma,H}=
-p_\mu\partial_\sigma x^\mu+p_\mu \tau^{\mu\nu}\tau_{\nu\rho}\partial_\sigma x^\rho
+B^\sigma=-p_\mu\partial_\sigma x^\mu+\Gamma^\sigma\approx -p_\mu\partial_\sigma x^\mu
\equiv -\mH_\sigma\approx 0 \ . 
\end{equation}
We see that the requirement of the preservation of the primary constraints
$P_\lambda^\tau\approx 0 \ , P_\lambda^\sigma\approx 0$  implies an existence of two secondary  constraints:
\begin{eqnarray}
\mH_\sigma=p_\mu\partial_\sigma x^\mu\approx 0 \ , \quad 
\mH_\tau=p_\mu h^{\mu\nu}p_\nu+\tau_F^2h_{\sigma\sigma}-2\tau_F p_\mu
\tau^\mu_{ \ a}\eta^{ab}e_{bc}\tau_\rho^{ \ c}\partial_\sigma x^\rho\approx 0 \ .
\nonumber \\
\end{eqnarray}
Further, since $\pb{P_B^\tau(\sigma),\Gamma^\tau(\sigma')}=\delta(\sigma-\sigma'), 
\pb{P_B^\sigma(\sigma),\Gamma^\sigma(\sigma')}=-\delta(\sigma-\sigma')$ we see that these constraints are the second class constraints that can be explicitly solved for $B^\tau$ and $B^\sigma$. Then these constraints vanish strongly and hence the 
Hamiltonian is linear combination of the constraints
\begin{equation}
\mH_E=\lambda^\tau \mH_\tau+\lambda^\sigma \mH_\sigma+v_\tau^\lambda P_\lambda^\tau+
v_\sigma^\lambda P_\lambda^\sigma  \ . 
\end{equation}
As the next step we should check that $\mH_\tau\approx 0 \ ,\mH_\sigma\approx 0$ 
are the first class constraints. To do this we introduce the smeared 
forms of these constraints 
\begin{eqnarray}
\bT_\tau(N)=\int d\sigma N \mH_\tau \ , 
\bT_\sigma(N^\sigma)=\int d\sigma N^\sigma \mH_\sigma \ . 
\nonumber \\
\end{eqnarray}
First of all we easily find 
\begin{eqnarray}
\pb{\bT_\sigma(N^\sigma),\bT_\sigma(M^\sigma)}=
%\int d\sigma (N^\sigma\partial_\sigma M^\sigma-N^\sigma 
%\partial_\sigma M^\sigma)p_\mu\partial_\sigma x^\mu=
\bT_\sigma(N^\sigma\partial_\sigma M^\sigma-N^\sigma 
\partial_\sigma M^\sigma) \ . \nonumber \\
\end{eqnarray}
In case of the Hamiltonian constraints the situation is more involved since the explicit calculation gives 
\begin{eqnarray}
& & \pb{\bT_\tau(N),\bT_\tau(M)}=\int d\sigma 
(N\partial_\sigma M-M\partial_\sigma N)4\tau_F^2p_\mu h^{
	\mu\nu}h_{\nu\rho}\partial_\sigma x^\rho
+
\nonumber \\
&+&2\int d\sigma \tau_F (N\partial_\sigma M-M\partial_\sigma N)
p_\mu V^\mu_{ \ \nu}h^{\nu\omega}p_\omega+
\nonumber \\
&+&\int d\sigma (N\partial_\sigma M-M\partial_\sigma N)
p_\rho V^\rho_{ \ \sigma}V^\sigma_{\ \omega}\partial_\sigma x^\omega+  
\nonumber \\
&+&\tau_F^2\int d\sigma (N\partial_\sigma M-M\partial_\sigma N)4\tau_F^2V^\mu_{ \ \nu}\partial_\sigma x^\nu h_{\mu\rho}\partial_\sigma x^\rho\ , 
\nonumber \\
\end{eqnarray}
where we defined 
\begin{equation}
V^\mu_{ \ \nu}=-2\tau_F\tau^\mu_{ \ a}\eta^{ab}\epsilon_{bc}
\tau_\nu^{\ c} \ . 
\end{equation}
To proceed further we calculate
\begin{eqnarray}
& &4p_\mu h^{\mu\nu}h_{\nu\rho}\partial_\sigma x^\rho=
4\tau_F^2p_\mu \partial_\sigma x^\mu-4\tau_F^2 p_\mu\tau^{\mu\rho}\tau_{\rho\nu}
\partial_\sigma x^\nu \ ,   \quad 
 \nonumber \\
& &p_\rho V^\rho_{ \ \mu}V^\mu_{ \ \nu}\partial_\sigma x^\nu=4\tau_F^2 p_\mu \tau^\mu_{ \ a}\tau^a_{ \ \nu}\partial_\sigma x^\nu=4\tau_F^2 p_\mu
\tau^{\mu\rho}\tau_{\rho\nu}\partial_\sigma x^\nu \ , \nonumber \\
& & V^\mu_{ \ \nu}\partial_\sigma x^\nu h_{\mu\rho}\partial_\sigma x^\rho=0 \ , 
\quad 
p_\mu V^\mu_{ \ \nu}h^{\nu\omega}p_\omega=0 \ .
\nonumber \\
\end{eqnarray}
Collecting all these results together we finally obtain  
\begin{equation}
\pb{\bT_\tau(N),\bT_\tau(M)}=\bT_\sigma ((N\partial_\sigma M-M\partial_\sigma N)
4\tau_F^2 ) \ . 
\end{equation}
Finally we also calculate  Poisson bracket between $\bT_\sigma(N^\sigma)$ and $\bT_\tau(M)$ and we find
\begin{equation}
\pb{\bT_\sigma(N^\sigma),\bT_\tau(M)}=
\bT_\tau(\partial_\sigma MN^\sigma-M\partial_\sigma N^\sigma) \ . 
\end{equation}
These results show that $\mH_\tau\approx 0,\mH_\sigma \approx 0 $ are the first class constraints and the non-relativistic string is well defined system from the Hamiltonian point of view.  

Finally we also say few words about the gauge fixed theory. We showed above that the Hamiltonian and spatial diffeomorphism constraints are the first class. Standard way how to deal with such a theory is to gauge fix them. For example, we can impose the static gauge when we introduce two gauge fixing functions 
\begin{equation}
\mG^0=x^0-\tau \approx 0 \ , \quad \mG^1=x^1-\sigma \approx 0 \ . 
\end{equation}
It is easy to see that 
 $\mG^a\approx 0$ are the second class constraints together with $\mH_\tau\approx 0 , \mH_\sigma\approx 0$. Since then these constraints vanish strongly we can identify the 
Hamiltonian density on the reduced phase space from the action principle
\begin{equation}
S=\int d\tau d\sigma (p_\mu \partial_\tau x^\mu-H)=
\int d\tau d\sigma (p_i\partial_\tau x^i+p_0)
\end{equation}
so that it is natural to identify $-p_0$ as the Hamiltonian on the reduced phase space 
$H_{red}=-p_0$. The explicit form of the Hamiltonian follows from the Hamiltonian constraint that can be solved for $p_0$.  Note also that $\mH_\sigma$ can be solved for $p_1$ as $p_1=-p_I\partial_\sigma x^I \ , I=2,\dots,d-1$.
\section{Generalization: Non-relativistic p-brane}\label{third}
In this section we perform generalization of the analysis presented 
previously to  the case of non-relativistic $p-$brane. 
As the first step we determine an action for non-relativistic p-brane in Newton-Cartan backgorund in the same way as in the string case. Explicitly, we start with the relativistic p-brane action coupled to $C^{p+1}$ form
whose action has the form 
\begin{equation}\label{pbraneaction}
S=-\ttau_p\int d^{p+1}\xi\sqrt{-\det \bA_{\alpha\beta}}
+\ttau_p\int C^{(p+1)} \ , \quad \bA_{\alpha\beta}=G_{\mu\nu}\partial_\alpha x^\mu 
\partial_\beta x^\nu \ , 
\end{equation}
where $\xi^\alpha \ , \alpha=0,\dots,p$ label world-volume of p-brane 
and where 
\begin{equation}
C^{(p+1)}=C_{\mu_1\dots \mu_{p+1}}dx^{\mu_1}\wedge \dots dx^{\mu_{p+1}}=
\frac{1}{(p+1)!}\epsilon^{\alpha_1\dots \alpha_{p+1}}
C_{\mu_1\dots \mu_{p+1}}\partial_{\alpha_1}x^{\mu_1}\dots \partial_{\alpha_{p+1}}x^{\mu_{p+1}} \ , 
\end{equation}
where again $\epsilon^{\alpha_1\dots \alpha_{p+1}}$ is totally antisymmetric
symbol. 

 With the help of the  action (\ref{pbraneaction}) we can proceed to the definition of non-relativistic p-brane in  Newton-Cartan background. As we have seen in case of the non-relativistic string
the requirement that the action for non-relativistic string should be finite
select two longitudial directions. Then we can deduce that in case of non-relativistic p-brane we should select $p+1$ longitudial directions. Explicitly, we presume that in case of the probe p-brane the background metric has the form 
\begin{eqnarray}
G_{\mu\nu}&=&E_\mu^{ \ a}E_\nu^{ \ b}\eta_{ab}+E_\mu^{ \ a'}E_\nu^{ \ b'}\delta_{a'b'}
=\nonumber \\
&=&\omega^2 \tau_{\mu\nu}+h_{\mu\nu}+\frac{1}{2}\tau_\mu^{ \ a}m_\nu^{ \ b}\eta_{ab}+
\frac{1}{2}m_\mu^{ \ a}\tau_\nu^{ \ b}\eta_{ab}+\frac{1}{4\omega^2}m_\mu^{ \ a}m_\nu^{ \ b}
\eta_{ab} \ ,  \nonumber \\ 
\end{eqnarray}
where now $a,b=0,\dots,p$ and $a',b'\dots=(p+1,\dots,d-1)$. Further, $\tau_{\mu\nu}$
and $h_{\mu\nu}$ are defined as 
\begin{equation}
\tau_{\mu\nu}=\tau_\mu^{ \ a}\tau_\nu^{ \ b}
\eta_{ab} \ , \quad \eta_{ab}=\mathrm{diag}(-1,\dots,1) \  , \quad h_{\mu\nu}=
e_\mu^{ \ a'}e_\nu^{ \ b'}\delta_{a'b'}  \ . 
\end{equation}
We also introduce their inverses with respect to their longitudinal and transverse dimensions
\begin{eqnarray}
e_\mu^{ \ a'}e^\mu_{ \ b'}=\delta^{a'}_{b'} \ ,  \quad
e_\mu^{ \ a'}e^\nu_{ \ a'}=\delta_\mu^\nu-\tau_\mu^{ \ a}
\tau^\nu_{ \ a} \ , \quad \tau^\mu_{ \ a}\tau_\mu^{ \ b}=\delta_a^b \ , \quad 
\tau^\mu_{ \ a}e_\mu^{ \ a'}=0 \ , \quad 
\tau_\mu^{ \ a}e^\mu_{ \ a'}=0 \ . \nonumber \\
\end{eqnarray}
In case of $p+1$-form $C^{(p+1)}$ we presume, with the analogy with the 
string case, that it has the form 
\begin{eqnarray}
C_{\mu_1\dots \mu_{p+1}}=\left(\omega\tau_{\mu_1}^{ \ a_1}-\frac{1}{2\omega}m_{\mu_1}^{ \ a_1}\right)\times \dots\times \left( \tau_{\mu_{p+1}}^{ \ a_{p+1}}-
\frac{1}{2\omega}m_{\mu_{p+1}}^{ \ a_{p+1}}\right)\epsilon_{a_1\dots a_{p+1}} \ , 
\end{eqnarray}
where  $\epsilon_{a_1\dots a_{p+1}}$ is  totally antisymmetric symbol. 
Now we are ready to define  non-relativistic limit of p-brane action. We start with the kinetic term and we obtain 
\begin{eqnarray}
S_{DBI}=-\ttau_p\omega^{p+1}\int d^{p+1}\xi\sqrt{-\det\ba}-
\frac{\ttau_p}{2}\omega^{p-1}\int d^{p+1}\xi\sqrt{-\det\ba}\tilde{\ba}^{\alpha\beta}
\bh_{\alpha\beta} \ ,
\nonumber \\
\end{eqnarray}
where $\tilde{\ba}^{\alpha\beta}$ is inverse to $\ba_{\alpha\beta}$. In fact, it is reasonable to presume that $\ba_{\alpha\beta}=\partial_\alpha x^\mu \eta_{ab}\partial_\beta x^\nu=\tau_\alpha^{ \ a}\eta_{ab}\tau_\beta^{ \ b}$ since 
$\tau_\alpha^{ \ a}$ and $\eta_{ab}$ are  $(p+1)\times (p+1)$ non-singular matrices.
From the requirement that the kinetic term is finite we have to perform following rescaling
\begin{equation}
\ttau_p \omega^{p-1}=\tau_p \ . 
\end{equation}
Further, the divergent term can be written as 
\begin{equation}
\ttau_p \omega^{p+1}\int d^{p+1}\xi \sqrt{-\det\ba}=
\tau_p \omega^2\int d^{p+1}\xi\det \tau_\alpha^{ \ b} \ , \quad   \tau_\alpha^{ \ a}=
\tau_\mu^{ \ a}\partial_\alpha x^\mu \ . 
\end{equation}
Let us now concentrate on the second term in the action 
(\ref{pbraneaction}). If we express $\ttau_p$ using $\tau_p$ as 
$\ttau_p=\frac{1}{\omega^{p-1}}\tau_p$ we find that the only non-zero
contribution comes from the product of $\tau_\mu^{ \ a}$'s while remaining terms vanish in the limit $\omega\rightarrow \infty$. Then we obtain 
\begin{eqnarray}
S_{WZ}&=&\frac{1}{\omega^{p-1}}\tau_p\int d^{p+1}\epsilon^{\alpha_1\dots \alpha_{p+1}}
\omega\tau_{\mu_1}^{ \ a_1}\partial_{\alpha_1}x^{\mu_1}\dots \omega\tau_{\mu_{p+1}}
^{ \ a_{p+1}}\partial_{\alpha_{p+1}}x^{\mu_{p+1}}\epsilon_{a_1\dots a_{p+1}}
\nonumber \\
&=&\omega^2\tau_p\int d^{p+1}\xi \frac{1}{p!}
\epsilon^{\alpha_1\dots \alpha_{p+1}}\tau_{\alpha_1}^{ \ a_1}
\dots \tau_{\alpha_{p+1}}^{ \ a_{p+1}}=
\omega^2\tau_p \int d^{p+1}\xi \det\tau_\alpha^{ \ a}
\nonumber \\
\end{eqnarray}
and we again see that these two divergent contributions cancel. As a result we obtain well defined action for non-relativistic p-brane in Newton-Cartan background   
\begin{equation}\label{nonrelpbraneaction}
S=-\frac{\tau_p}{2}\int d^{p+1}\xi 
\sqrt{-\det\ba}\tilde{\ba}^{\alpha\beta}\bh_{\mu\nu}\partial_\alpha x^\mu
\partial_\beta x^\nu \ .
\end{equation}
Now we proceed to the Hamiltonian formulation of this theory. With the analogy with the string case we write the action as 
\begin{eqnarray}\label{pbraneactionextended}
S&=&\int d^{p+1}
\xi \left(\frac{1}{4\lambda^\tau}(\partial_0 x^\mu-\lambda^i\partial_i x^\mu)
h_{\mu\nu}(\partial_0 x^\nu-\lambda^j\partial_j x^\nu)-\lambda^\tau \tau^2_p\det\ba_{ij}\ba^{ij}h_{ij} \right. 
\nonumber \\
& &\left.+B^0\left(\lambda^0-\frac{\sqrt{-\det \ba}}{2\tau_p
\det \ba_{ij}}\right)+B^i(\lambda_i-\ba_{i0})\right) \ , 
\nonumber \\
\end{eqnarray}
where
\begin{equation}
\lambda^i=\ba^{ij}\ba_{j0} \ ,  \quad 
\ba_{ij}\ba^{jk}=\delta_i^ k \ . 
\end{equation}
In order to see
 an equivalence between the action (\ref{pbraneactionextended}) and 
(\ref{nonrelpbraneaction})  we note that the inverse matrix $\tilde{\ba}^{\alpha\beta}$ to the matrix $\ba_{\alpha\beta}$ has  the form 
\begin{eqnarray}\label{inversematrix}
\tilde{\ba}^{00}&=&\frac{\det\ba_{ij}}{\det\ba} \ , \quad 
\tilde{\ba}^{0i}=-\ba_{0k}\ba^{kj}\frac{\det\ba_{ij}}{\det\ba} \ ,  \nonumber \\
\tilde{\ba}^{i0}&=&-\ba^{ik}\ba_{k0}\frac{\det\ba_{ij}}{\det\ba} \ , \quad 
\tilde{\ba}^{ij}=\ba^{ij}+\frac{\det\ba_{ij}}{\det\ba}
\ba^{ik}\ba_{k0}\ba_{0l}\ba^{lj} \ , \nonumber \\ 
\end{eqnarray}
where $\ba^{ij}\ba_{jk}=\delta^i_k$. 
 Then the equation of motion for 
 $B^0$ and $B^i$ imply
\begin{equation}
\lambda^0=-\frac{\sqrt{-\det\ba}}{2\tau_p \det\ba_{ij}} \ , \quad  \lambda_i=\ba_{0i} \ . 
\end{equation}
Inserting this result into (\ref{pbraneactionextended}) we obtain that it is equal to the action (\ref{nonrelpbraneaction}).\
%\begin{eqnarray}
%\mL
%\nonumber%=\frac{1}{4\lambda^\tau }h_{00}-\frac{\lambda^i}{2\lambda^\tau}h_{0i}+
%%\frac{1}{4\lambda^\tau}(\lambda^i \lambda^j-4\tau_p^2(\lambda^\tau)^2\sqrt{\det\ba_{ij}}\ba^{ij})h_{ij}=
%=-\frac{1}{2\tau_p}\sqrt{-\det\ba}(\ba^{00}h_{00}
%-\frac{2\ba^{ij}\ba_{j0}\det\ba_{ij}}{\det\ba}h_{0i})+
%\frac{1}{4\lambda^\tau}\frac{\det\ba}{\det\ba_{ij}}
%(\ba_{0k}\ba^{ki}\ba_{0l}\ba^{lj}\frac{\det\ba_{ij}}{\det \ba}+
%\ba^{ij})h_{ij}=\nonumber \\
%=-\frac{1}{2\tau_p}\sqrt{-\det\ba}(\ba^{00}h_{00}+2\ba^{0i}h_{0i}+\ba^{ij}h_{ij})
%\nonumber \\
%\end{eqnarray}
Let us now return to the action  (\ref{pbraneactionextended})
and determine conjugate momenta from it 
\begin{eqnarray}\label{pmubrane}
& &p_\mu=\frac{\partial \mL}{\partial (\partial_0 x^\mu)}\nonumber \\
&=&\frac{1}{2\lambda^\tau}\bh_{\mu\nu}\partial_0 x^\nu-\frac{\lambda^i}{2\lambda^\tau}
\bh_{\mu\nu}\partial_i x^\nu-B^\tau\frac{1}{2\tau_p\det\ba_{ij}}
\tau_{\mu\nu}\partial_\alpha x^\nu \bai^{\alpha 0}\sqrt{-\det\ba}-
B^i\tau_{\mu\nu}\partial_i x^\nu \ , \nonumber \\
\nonumber \\
& &P_0=\frac{\partial \mL}{\partial(\partial_0 \lambda^0)}\approx 0 \ , \quad 
P_i=\frac{\partial \mL}{\partial(\partial_0 \lambda^i)}\approx 0 \ , \quad 
P^B_0=\frac{\partial \mL}{\partial(\partial_0 B^0)}\approx 0 \ , \quad 
P^B_i=\frac{\partial \mL}{\partial(\partial_0 B^i)}\approx 0 \ . 
\nonumber \\
\end{eqnarray}
From the same reason as in case of the fundamental string we have to restrict
to the case $m_\mu^{ \ a}=0$ so that $\bh_{\mu\nu}=h_{\mu\nu}$. Then if we multiply 
(\ref{pmubrane}) with  $h^{\mu\nu}$  we obtain
\begin{eqnarray}
%h^{\mu\nu}(p_\nu+\frac{\lambda^i}{2\lambda^\tau}h_{\mu\nu}\partial_i x^\nu)=
%\frac{1}{2\lambda^\tau}h_{\mu\nu}\partial_0 x^\nu 
%\nonumber \\
\left(p_\mu+\frac{\lambda^i}{2\lambda^\tau}h_{\mu\rho}\partial_i x^\rho\right)
h^{\mu\nu}\left(p_\nu+\frac{\lambda^j}{2\lambda^\tau}h_{\nu\sigma}\partial_jx^\sigma\right)=
\frac{1}{4(\lambda^\tau)^2}\partial_0 x^\mu h_{\mu\nu}\partial_0 x^\nu \ . \nonumber \\
\end{eqnarray}
On the other hand let us multiply both sides of (\ref{pmubrane}) with $\tau^{\mu\nu}\tau_{\nu\rho}
\partial_i x^\rho$ and  we obtain 
\begin{eqnarray}
p_\mu \tau^{\mu\nu}\tau_{\nu\rho}\partial_i x^\rho=
-B^\tau \frac{1}{2\tau_p\det\ba_{ij}}
\partial_i x^\mu \tau_{\mu\nu}\partial_\alpha x^\nu \ba^{\alpha 0}
\sqrt{-\det\ba}-\partial_i x^\mu \tau_{\mu\nu}\partial_j x^\nu B^j=
-\ba_{ij}B^j 
\nonumber \\
\end{eqnarray}
that imlies $p-$primary constraints
\begin{eqnarray}\label{primpcon}
\Sigma^i\equiv p_\mu \tau^{\mu\nu}\tau_{\nu\sigma}\partial_j x^\sigma \ba^{ji}+B^i \approx 0  \ . 
\nonumber \\
\end{eqnarray}
On the other hand let us multiply (\ref{pmubrane}) with following expression 
\begin{equation}
\frac{1}{p!}\tau^\mu_{ \ a}\eta^{aa_1}\epsilon_{a_1\dots a_{p+1}}
\tau^{\ a_2}_{\nu_2}\dots \tau^{\ a_{p+1}}_{\nu_{p+1}}
\epsilon^{j_2\dots j_{p+1}}\partial_{j_2}x^{\nu_2}\dots \partial_{j_{p+1}}
x^{\nu_{p+1}} \ . 
\end{equation}
Then using the fact that 
\begin{eqnarray}
& &\partial_i x^\nu\tau_{\nu\mu}\tau^\mu_{ \ a}\eta^{aa_1}
\epsilon_{a_1\dots a_{p+1}}\tau_{\nu_2}^{ \ a_2}
\dots \tau_{\nu_{p+1}}^{\ a_{p+1}}\epsilon^{j_2\dots j_{p+1}}
\partial_{j_2}x^{\nu_2}\dots \partial_{j_{p+1}}x^{\nu_{p+1}}=0  \ , 
\nonumber \\
%=\partial_i x^\nu\tau_{\nu}^{ \ a_1}\epsilon_{a_1\dots a_{p+1}}
%\tau_{\nu_2}^{ \ a_2}\dots \tau_{\nu_{p+1}}^{ \ a_{p+1}}
%\epsilon^{j_2\dots j_{p+1}}
%\partial_{j_2}x^{\nu_2}\dots \partial_{j_{p+1}}x^{\nu_{p+1}}=0
%\nonumber \\
%\end{eqnarray}
%and
%\begin{eqnarray}
& &\frac{1}{p!}\ba^{0\alpha}\partial_\alpha x^\nu \tau_{\nu\mu}
\tau^\mu_{ \ a}\eta^{aa_1}\epsilon_{a_1\dots a_{p+1}}
\tau_{\nu_2}^{ \ a_2}\dots \tau_{\nu_{p+1}}^{ \ a_{p+1}}
\epsilon^{j_2\dots j_{p+1}}\partial_{j_2}x^{\nu_2}\dots \partial_{j_{p+1}}
x^{\nu_{p+1}}\frac{\sqrt{-\det\ba}}{\det\ba_{ij}}=\nonumber \\
%\frac{1}{p!}\ba^{0\alpha}\partial_\alpha x^\nu \tau_\nu^{ \ a_1}
%\epsilon_{a_1\dots a_{p+1}}
%\tau_{\nu_2}^{ \ a_2}\dots \tau_{\nu_{p+1}}^{ \ a_{p+1}}
%\epsilon^{j_2\dots j_{p+1}}\partial_{j_2}x^{\nu_2}\dots \partial_{j_{p+1}}
%x^{\nu_{p+1}}\frac{\sqrt{-\det\ba}}{\det\ba_{ij}}=\nonumber \\
%=\frac{1}{p!}\ba^{00}\partial_0 x^\nu \tau_\nu^{ \ a_1}
%\epsilon_{a_1\dots a_{p+1}}
%\tau_{\nu_2}^{ \ a_2}\dots \tau_{\nu_{p+1}}^{ \ a_{p+1}}
%\epsilon^{j_2\dots j_{p+1}}\partial_{j_2}x^{\nu_2}\dots \partial_{j_{p+1}}
%x^{\nu_{p+1}}\frac{\sqrt{-\det\ba}}{\det\ba_{ij}}=
%\nonumber \\
%=-\frac{1}{p!}\partial_0 x^\nu \tau_\nu^{ \ a_1}
%\epsilon_{a_1\dots a_{p+1}}
%\tau_{\nu_2}^{ \ a_2}\dots \tau_{\nu_{p+1}}^{ \ a_{p+1}}
%\epsilon^{j_2\dots j_{p+1}}\partial_{j_2}x^{\nu_2}\dots \partial_{j_{p+1}}
%x^{\nu_{p+1}}\frac{1}{\sqrt{-\det\ba}}=
%\nonumber \\
& &=-\frac{1}{(p+1)!}\epsilon_{a_1\dots a_{p+1}}
\epsilon^{j_1\dots j_{p+1}}\tau_{\nu_1}^{ \ a_1}\partial_{\alpha_1}x^{\nu_1}
\dots \tau_{\nu_{p+1}}^{ \ a_{p+1}}\partial_{\alpha_{p+1}}x^{\nu_{p+1}}
\frac{1}{\det \tau_\alpha^{ \ a}}
=-1  \nonumber \\
\end{eqnarray}
 we obtain second primary constraint
\begin{equation}
\Sigma^0 \equiv 
2\tau_p p_\mu \frac{1}{p!}\tau^\mu_{ \ a}\eta^{aa_1}\epsilon_{a_1\dots a_{p+1}}
\tau^{\ a_2}_{\nu_2}\dots \tau^{\ a_{p+1}}_{\nu_{p+1}}
\epsilon^{j_2\dots j_{p+1}}\partial_{j_2}x^{\nu_2}\dots \partial_{j_{p+1}}
x^{\nu_{p+1}}-B^0 \approx 0 \ . 
\end{equation}
%As the next step we determine bare Hamiltonian 
%\begin{eqnarray}
%H_B&=&\int d^p\xi (p_\mu\partial_0 x^\mu-\mL)=\nonumber \\
%%=\int d^p\xi(\frac{1}{4\lambda^\tau }\partial_0 x^\mu h_{\mu\nu}\partial_0 x^\nu+\lambda^\tau \tau_p^2\sqrt{\det\ba_{ij}}\ba^{ij}h_{ij}
%%-\frac{1}{4\lambda^\tau}\lambda^i\lambda^j h_{ij}-B^0\lambda^0-B^i\lambda_i)=
%%\nonumber \\
%&=&\int d^p\xi (\lambda^0 p_\mu h^{\mu\nu}p_\nu+\lambda^i p_\mu h^{\mu\nu}h_{\nu\sigma}
%\partial_i x^\sigma+\lambda^\tau \tau_p^2\det\ba_{ij}\ba^{ij}h_{ij}
%-B^0\lambda^0-B^i\lambda^i)
%\nonumber \\
%\end{eqnarray}
Using all these results we determine extended Hamiltonian with all primary constraints included in the form 
\begin{eqnarray}\label{Hamextendedbrane}
H_E=\int d^p\xi (\lambda^0 p_\mu h^{\mu\nu}p_\nu+\lambda^i p_\mu h^{\mu\nu}h_{\nu\sigma}
\partial_i x^\sigma+\lambda^\tau \tau_p^2\det\ba_{ij}\ba^{ij}h_{ij}-
\nonumber \\
-B^0\lambda^\tau-B^i\lambda_i+v^0P_0+v^iP_i+v^0_BP_0^B+v_i^B P^i_B
+\Psi_0\Sigma^0+\Psi_i\Sigma^i) \ . 
\nonumber \\
\end{eqnarray}
Since $\pb{P_0^B(\xi),\Sigma^0(\xi')}=\delta(\xi-\xi') \ , 
\pb{P^i_B(\xi),\Sigma^j(\xi')}=-\delta^{ij}\delta(\xi-\xi')$ we see that 
 that $P_0^B$ together with $\Psi^0$ are the couple of $p+1$ second class constraints. Then we can solve these constraints with respect to $B^0,B^i$ and we 
we obtain the Hamiltonian 
in the form 
\begin{eqnarray}
H_B
%=\int d^p\xi (\lambda^0(p_\mu h^{\mu\nu}p_\nu+
%\tau_p^2\det\ba_{ij}\ba^{ij}h_{ij})+
%\lambda^i p_\mu \partial_\sigma x^\mu
%-\lambda^i p_\mu \tau^{\mu\nu}\tau_{\nu\rho}\partial_\sigma x^\rho
%+\nonumber \\
%+p_\mu \tau^{\mu\nu}\tau_{\nu\rho}\partial_j x^\nu 
%\ba^{ji}\lambda_i-\nonumber \\
%-\lambda^0
%2\tau_p p_\mu \frac{1}{p!}\tau^\mu_{ \ a}\eta^{aa_1}\epsilon_{a_1\dots a_{p+1}}
%\tau^{\ a_2}_{\nu_2}\dots \tau^{\ a_{p+1}}_{\nu_{p+1}}
%\epsilon^{j_2\dots j_{p+1}}\partial_{j_2}x^{\nu_2}\dots \partial_{j_{p+1}}
%x^{\nu_{p+1}}+v^0P_0+v^iP_i)=\nonumber \\
=\int d^p\xi (\lambda^0\mH_0+\lambda^i\mH_i+v^0P_0+v^i P_i) \nonumber \\
\end{eqnarray}
where 
\begin{eqnarray}
\mH_0&=&p_\mu h^{\mu\nu}p_\nu
+\tau_p^2\det\ba_{ij}\ba^{ij}h_{ij} -\nonumber \\
&-&2\tau_p p_\mu \frac{1}{p!}\tau^\mu_{ \ a}\eta^{aa_1}\epsilon_{a_1\dots a_{p+1}}
\tau^{\ a_2}_{\nu_2}\dots \tau^{\ a_{p+1}}_{\nu_{p+1}}
\epsilon^{j_2\dots j_{p+1}}\partial_{j_2}x^{\nu_2}\dots \partial_{j_{p+1}}
x^{\nu_{p+1}}\approx 0 \ ,  \nonumber \\
\mH_i&=&p_\mu\partial_ix^\mu \approx 0 \ . 
\nonumber \\
\end{eqnarray}
Then the requirement of the preservation of the constraint $P_0\approx 0\ ,  P_i\approx 0$ implies $p+1$ secondary constraints
\begin{equation}
\mH_0\approx 0 \ , \quad \mH_i\approx 0 \ . 
\end{equation}
Now we have to check that these constraints are the first class constraints. We 
introduce their smeared form 
\begin{equation}
\bT_T(N)=\int d^p\xi N\mH_0 \ , \quad 
\bT_S(N^i)=\int d^p\xi N^i\mH_i \ 
\end{equation}
and calculate corresponding  Poisson brackets. First of all we have
\begin{eqnarray}
\pb{\bT_S(N^i),\bT_S(M^j)}=\bT_S(N^j\partial_j M^i-M^j\partial_j N^i) \ . 
 \nonumber \\
\end{eqnarray}
In case of the calculation of the Poisson brackets between $\bT_S(N^i)$ and 
 $\bT_T(M)$ we have
to be more careful. First of all we have
\begin{equation}
\pb{\bT_S(N^ i),\tau_i^{ \ a}}=-N^k\partial_k \tau_i^{ \ a}-
\partial_i N^j \tau_j^{ \ a} \ , \quad \tau_i^{ \ a}\equiv \partial_i x^\mu
\tau_\mu^{ \ a} \ . 
\end{equation}
Then we obtain 
\begin{eqnarray}
& &\pb{\bT_S(N^i),\ba_{ij}}=-N^k\partial_k \ba_{ij}-\partial_i N^k\ba_{kj}-
\ba_{ik}\partial_j N^k \ ,  \nonumber \\
& &\pb{\bT_S(N^i),\ba^{ij}}=-N^k\partial_k \ba^{ij}+\partial_k N^i\ba^{kj}+
\ba^{ik}\partial_k N^j \ , \nonumber \\
& &\pb{\bT_S(N^i),\det\ba_{ij}}=-N^k\partial_k (\det\ba_{ij})-
2\partial_i N^i\det\ba_{ij} \ . \nonumber \\
\end{eqnarray}
Using also the fact that 
\begin{eqnarray}
& &\pb{\bT_S(N^i),\partial_i x^\mu}=-N^k\partial_k(\partial_i x^\mu)-\partial_i
N^k\partial_k x^\mu \ , \nonumber \\
& &\pb{\bT_S(N^i),h_{ij}}=-N^k\partial_k h_{ij}-\partial_i N^k h_{kj}
-h_{ik}\partial_j N^k \ \nonumber \\
\end{eqnarray}
we finally obtain 
\begin{equation}
\pb{\bT_S(N^i),\det\ba_{ij}\ba^{kl}h_{kl}}=-N^k\partial_k (\det\ba_{ij}
\ba^{kl}h_{kl})-2\partial_i N^i\det\ba_{ij}\ba^{kl}h_{kl} \ . 
\end{equation}
Let us introduce following vector 
\begin{equation}
V^\mu=-2\tau_p \frac{1}{p!}\tau^\mu_{ \ a}\eta^{aa_1}\epsilon_{a_1\dots a_{p+1}}
\tau^{\ a_2}_{\nu_2}\dots \tau^{\ a_{p+1}}_{\nu_{p+1}}
\epsilon^{j_2\dots j_{p+1}}\partial_{j_2}x^{\nu_2}\dots \partial_{j_{p+1}}
x^{\nu_{p+1}} \ . 
\end{equation}
Then after some algebra we obtain 
\begin{equation}
\pb{\bT_S(N^i),V^\mu}=-N^k\partial_k V^\mu-2\partial_k N^k V^\mu \ . 
\end{equation}
Collecting all these results together we finally find 
%and finally we obtain 
%\begin{eqnarray}
%\pb{\bT_S(N^i),\tau_p\frac{p_\mu}{p!}
%\tau^\mu_{ \ a}\eta^{aa_1}\epsilon_{a_1\dots a_{p+1}}
%\tau_{\nu_2}^{ \ a_2}\dots \tau_{\nu_{p+1}}^{ \ a_{p+1}}
%\epsilon^{j_2\dots j_{p+1}}\partial_{j_2}x^{\nu_2}\dots \partial_{j_{p+1}}x^{\nu_{p+1}}
%}=\nonumber \\
%=- \partial_i N^ i \tau_p\frac{p_\mu}{p!}
%\tau^\mu_{ \ a}\eta^{aa_1}\epsilon_{a_1\dots a_{p+1}}
%\tau_{\nu_2}^{ \ a_2}\dots \tau_{\nu_{p+1}}^{ \ a_{p+1}}
%\epsilon^{j_2\dots j_{p+1}}\partial_{j_2}x^{\nu_2}\dots \partial_{j_{p+1}}x^{\nu_{p+1}}
%-\nonumber \\
%-N^i\partial_i[\tau_p\frac{p_\mu}{p!}
%\tau^\mu_{ \ a}\eta^{aa_1}\epsilon_{a_1\dots a_{p+1}}
%\tau_{\nu_2}^{ \ a_2}\dots \tau_{\nu_{p+1}}^{ \ a_{p+1}}
%\epsilon^{j_2\dots j_{p+1}}\partial_{j_2}x^{\nu_2}\dots \partial_{j_{p+1}}x^{\nu_{p+1}}
%]-\nonumber \\
%-\tau_p\frac{p_\mu}{p!}
%\tau^\mu_{ \ a}\eta^{aa_1}\epsilon_{a_1\dots a_{p+1}}
%\partial_{i_1}N^= k\tau_{\nu_2}^{ \ a_2}\dots \tau_{\nu_{p+1}}^{ \ a_{p+1}}
%\epsilon^{j_2\dots j_{p+1}}\partial_{k}x^{\nu_2}\dots \partial_{j_{p+1}}x^{\nu_{p+1}}-
%\nonumber \\
%-\dots
%\tau_p\frac{p_\mu}{p!}
%\tau^\mu_{ \ a}\eta^{aa_1}\epsilon_{a_1\dots a_{p+1}}
%\tau_{\nu_2}^{ \ a_2}\dots \tau_{\nu_{p+1}}^{ \ a_{p+1}}
%\epsilon^{j_2\dots j_{p+1}}\partial_{j_2}x^{\nu_2}\dots \partial_{j_{p+1}}N^k
%\partial_kx^{\nu_{p+1}}
%\nonumber \\
%\end{eqnarray}
\begin{equation}
\pb{\bT_S(N^i),\bT_S(M)}=\bT_T(N^i\partial_iM-\partial_i N^i M) \ . 
\end{equation}
Finally we calculate Poisson brackets of the smeared forms of Hamiltonian 
constraints
and we obtain 
\begin{eqnarray}\label{pbTTpbrane}
& &\pb{\bT_T(N),\bT_T(M)}=\int d^p\xi (N\partial_iM-M\partial_iN)4\tau_p^2
\det\ba_{ij}\ba^{ij}p_\mu h^{\mu\nu}h_{\nu\sigma}\partial_j x^\sigma+
\nonumber \\
&+&2\tau_p\int d^p\xi
(N\partial_i M-M\partial_iN)\frac{1}{(p-1)!}p_\nu\tau^\nu_{ \ a}\eta^{aa_1}
\epsilon_{a_1\dots a_{p+1}}\tau_{\mu}^{ \ a_2}\tau_{\nu_3}^{ \ a_3}
\dots \tau_{\nu_{p+1}}^{ \  a_{p+1}}\times \nonumber \\
&\times & 
\epsilon^{i j_3\dots j_{p+1}}\partial_{j_3}x^{\nu_3}\dots \partial_{j_{p+1}}
x^{\nu_{p+1}}V^\mu \ .  \nonumber \\  
\end{eqnarray}
Then after some lengthy calculations we find 
that the last expression is equal to 
\begin{equation}
4\tau_p^2 (N\partial_i M-M\partial_i N)\ba^{ij}p_\mu \tau^{\mu\nu}
\tau_{\nu\sigma}\partial_j x^\sigma \det \ba_{ij} \ . 
\end{equation}
Inserting this result into (\ref{pbTTpbrane} we obtain  final result
\begin{equation}
\pb{\bT_T(N),\bT_T(M)}=\bT_S((N\partial_iM-M\partial_iN)4\tau_p^2\ba^{ij} 
\det\ba_{ij}) \ . 
\end{equation}
These results show that $\mH_0$ and $\mH_i$ are the first class constraints 
that reflect diffeomorphism invariance of non-relativistic p-brane.

\section{Conclusion}\label{fourth}
In this paper we formulated non-relativistic actions for string and p-brane in Newton-Cartan background. Then we found their Hamiltonian formulations and we determined structure of constraints in the special case where the gauge field $m_\mu^{ \ a}$ is zero. We argued that we restricted to this case since we were not able to express time derivatives of $x^\mu$ or their combinations as functions of canonical variables in the case when $m_\mu^{ \ a}\neq 0$. Certainly this more general case deserves further study.
One possibility is to  address this problem from different point of view when we
start with the Hamiltonian formulation string in general background, then we 
perform limiting procedure on the background metric and NSNS two form field 
and derive corresponding Hamiltonian. This problem is currently under study and we 
hope to report on this analysis in near future.

\acknowledgments{This  work  was
	supported by the Grant Agency of the Czech Republic under the grant
	P201/12/G028. }

%%%%%%%%%%%%%%%%%%%%%%%%%%%%%%%%%%%%%%
%%%%%%% Thebibligraphy %%%%%%%%%%%
%%%%%%%%%%
%%%%%%%%%%%%%%%%%%%%%%%%%%%%%%%%%%%%%

\end{document}